\documentstyle[12pt,a4wide]{article}
\input{epsf}
\begin{document}
\setlength\textheight{8.75in}
\setlength\textheight{9.00in}
%%%%%%%%%%%%
\newcommand{\be}{\begin{equation}}
\newcommand{\ee}{\end{equation}}
\title{Quasi exactly solvable quantum lattice solitons}
\author{
{\large Yves Brihaye\footnote{ yves.brihaye@umh.ac.be}} \\
{\small Facult\'e des Sciences, Universit\'e de Mons-Hainaut, }\\
{\small B-7000 Mons, Belgium }\\
{ } \\
 {\large Nathalie Debergh
   \footnote{n.debergh@ulg.ac.be}}\\
{\small   Techniques du Son et de l'Image, Mont\'efiore Institute,
Universit\'e de Li\`ege}\\
{\small B-4000 Li\`ege, Belgium}
{ } \vspace{5mm} \\
   {\large Ancilla Nininahazwe
   \footnote{nininaha@yahoo.fr}}\\
{\small   Facult\'e des Sciences,
Universit\'e du Burundi}\\
{\small P.O. Box 2700, Bujumbura, Burundi}\\ }

\date{\today}
%%%%%%\begin{titlepage}
\maketitle
\thispagestyle{empty}
\begin{abstract}
We extend the exactly solvable Hamiltonian describing
$f$ quantum oscillators considered recently by
J. Dorignac et al. by means of a new interaction
which we choose as quasi exactly solvable. The properties of the
spectrum of this new Hamiltonian are studied as function of the
new coupling constant. This Hamiltonian as well as the original 
one are also related to adequate Lie structures.
\end{abstract}
\medskip
\medskip
\newpage
\section{Introduction}
A few years ago the Bose-Hubbard (BH) model describing a
non linear optic model
with  $f$ bosonic oscillators in interaction was studied in some
details \cite{scott},\cite{eilbeck},\cite{eilbeck1}. In particular
 it was revealed that the spectrum of the restriction
of the Hamiltonian to the subspace of vectors involving two quanta
has a remarkable property when the limit of large $f$ is considered~:
the spectrum separates into two pieces, one line of discrete states
forming the so called "soliton band" and another region where
the eigenvalues form a continuum. The total Hamiltonian
describing the BH model
commutes with the operator $N$ counting the number of quanta.
As a consequence,
the Fock space ${\cal V}$ of quantum states can be decomposed
into an infinite flag of finite-dimensional
subspaces  ${\cal V}_n$, $n=0,1,2,\dots$ that are left invariant
by the Hamiltonian and the whole spectrum can be constructed
algebraically. Recently, these ideas were generalized to exactly solvable
Hamiltonians with higher orders in the particle operators
and it was shown that similar results hold.
\par
Such properties are intimately connected to specific features of Lie structures. More precisely, we prove in Section 2 that the BH model is the sum of scaling operators of the Lie algebra $sl(f; R)$ supplemented by bilinear fonctions of the (diagonal) operators generating the Cartan subalgebra of this structure. Thus, the number $f$ of sites fixes the Lie structure subtended by the model. We also prove the fundamental role played by the number $n$ of quanta picking up the irrreducible representations (irreps) that are concerned with.

Operators enjoying the algebraic  property above are called
exactly solvable \cite{turbiner1}.
The family of exactly solvable Hamiltonians is rather small;
however if we replace the requirement that the Hamiltonian
preserves an infinite flag of finite-dimensional subspaces by the
weaker requirement that one finite dimensional subspace is preserved
by the Hamiltonian, we are left with the notion of 
quasi-exactly-solvable (QES) \cite{turbiner}
operators and/or equations.
In this case not all but a
{\it finite} part of the spectrum can be computed algebraically.

In Section 3, we apply the ideas of QES equations and we extend
the exactly solvable BH Hamiltonian by a new term which
preserves only a finite-dimensional subspace of the Fock space 
${\cal V}$ containing  the subspace ${\cal V}_2$ where
the splitting of the eigenvalues occurs in the normal 
BH model. The new term is characterized by a new coupling
constant, say $\lambda$.
\par
Then, in Section 4, we relate the 
operators involved in this new model to 
generators of an ad-hoc Lie structure namely the 
Lie orthosymplectic superalgebra $osp(1 / 2f ; R)$.

 Section 5 is devoted to the spectrum of this QES operator
in function of $\lambda$. The analysis
of the eigenvalues can be achieved along the same lines as in 
\cite{scott} and our results suggest that the splitting between
the soliton band and the continuum still occurs for the 
QES Hamitonian. We illustrate these results by some examples in Section 6.
\section{Group theoretical approach of the BH model}
The BH Hamiltonian is \cite{scott,eilbeck}
\be
H_{BH} = -\sum_{j=1}^f[a_j^{\dagger}a_{j+1} + a_j^{\dagger}a_{j-1}
+ \frac{\gamma}{2}a_j^{\dagger}a_j^{\dagger}a_ja_j]
\ee
where the bosonic lowering and raising operators $a_j ,a_j^{\dagger}$
obey the usual commutation rules
$[a_i,a_j]=[a_i^{\dagger},a_j^{\dagger} ]=0$, 
$[a_i,a_j^{\dagger} ]= \delta_{ij}$
and  the following periodic conditions~:
\be
a_{f+1}=a_1 \ , \ a_{f+1}^{\dagger} = a_1^{\dagger}.
\ee
The main property of the operator $H_{BH}$ is that
it preserves separately any subspace ${\cal V}_n$ of the Hilbert space
with $n$ quanta, i.e. the space generated by the vectors
of the form $\vert n_1,n_2,\dots, n_f \rangle$ for fixed
$n = n_1+n_2+\dots +n_f$.
\par
Let us now consider the group theoretical approach of such a model. To do so, we fix the grading of the operator $a^{\dagger}_j a_k$ as $(j-k)$. The interacting part of the BH Hamiltonian is thus a linear combination of $(f-1)$ operators of grading $+1$ and $(f-1)$ other operators of grading $-1$. Following the usual Lie bracket as well as the grading rule
\be
[ grading(j) , grading(k)  ] = grading(j+k),
\ee
the operators of grading $+1$ i.e. $a_j^{\dagger}a_{j-1}$ give rise to $(f-2)$ operators of grading $+2$ namely $a_j^{\dagger}a_{j-2}$ while the operators of grading $-1$ i.e. $a_j^{\dagger}a_{j+1}$ generate $(f-2)$ operators of grading $-2$ given by $a_j^{\dagger}a_{j+2}$. The process goes on and on until the operators $a^{\dagger}_{f} a_1$ and $a^{\dagger}_{1} a_f$ of respective gradings $(f-1)$ and $(1-f)$ are reached. To these scaling operators, we add the $(f-1)$ diagonal (or of grading $0$) ones $a^{\dagger}_{j+1} a_{j+1}-a^{\dagger}_{j} a_j$ with $j = 1, ..., f - 1$ and we finally obtain $2 ( \sum_{j=1}^{f-1} j )+ f - 1 = f^2 - 1$ operators generating the Lie algebra $sl( f ; R )$.
\par
Let us now turn to some specific examples.
\par
The case $f=2$ is the first significant one. The Lie algebra $sl(2 ; R)$ is the one subtended by the BH model. It is characterized by the commutation relations
\be
[ J_0, J_{\pm} ] = \pm 2 J_{\pm} \; , \; [J_+ , J_- ] = J_0
\ee
and the three generators are realized through
\be
J_0 = a^{\dagger}_{2} a_{2}-a^{\dagger}_{1} a_1 \; , \; J_+ = a^{\dagger}_{2} a_{1} \; , \; J_- = a^{\dagger}_{1} a_{2}.
\ee
More precisely, a rapid look at the Casimir operator
\be
 C = J_+ J_- + \frac{1}{4} J_0^2 - \frac{1}{2} J_0
\ee
can convince us that
\be
C | n_1 , n_2 > = \frac{1}{4} n (n+2)  | n_1 , n_2 >
\ee
or in other words that the so-called $D^{(\frac{n}{2})}$ irrep of $sl(2 ; R)$ is under consideration in the BH model. 
\par
Such a representation has the feature that it can be realized through the following differential operators 
\be
J_0= 2 x \frac{d}{dx} - n \; , \; J_+ = -x^2 \frac{d}{dx} + n x \; , \; J_- = \frac{d}{dx}
\ee
while the basis state $| n-n_2 , n_2 >$ is assimilated to the monomial $x^{n_2}$ for $n_2 = 0, 1, ..., n$. The BH Hamiltonian (1) is then
\begin{eqnarray}
H_{BH} & = & - J_+ - J_- + \frac{\gamma}{4} (2N - N^2 - J_0^2) \\
& = & - \gamma x^2 \frac{d^2}{dx^2} + [ (x^2-1) + \gamma (n-1) x ] \frac{d}{dx} - n x - \frac{\gamma}{2} n (n-1).
\end{eqnarray}
where $N$ is the number operator of eigenvalue $n$.
It can be put on a Schr\" odinger form if the change of variables
\be
x = \exp{(\pm \sqrt{\gamma}y)}
\ee
as well as the "Gauge transformation" 
\be
\chi =  \exp{(\pm \frac{n}{2} \sqrt{\gamma}y)} \exp{(\frac{1}{\gamma} \cosh{\sqrt{\gamma}y})} \psi
\ee
are performed. The resulting potential writes
\be
V (y) = \frac{1}{\gamma} \cosh^2{\sqrt{\gamma}y}-(n+1) \cosh{\sqrt{\gamma}y}-\frac{1}{\gamma} - \gamma \frac{n}{2} (\frac{n}{2} -1).
\ee
It coincides with the one studied in \cite{DVDB} (cf. Eq. (170)) and we refer to this paper for the determination of the corresponding eigenvalues and eigenfunctions.
\par
An operator playing a fundamental role in this knowledge of the eigenvalues and eigenfunctions is the translation operator $T$ \cite{scott,eilbeck}. It is defined by the property $Ta_j^{\dagger} = a_{j+1}^{\dagger}T$  so that $T|n_1,n_2,\dots ,n_f>= |n_f, n_1,n_2,\dots , n_{f-1}>$ . The BH Hamiltonian as constructed in Eq. (1)  commutes with $T$. 
\par
Let us consider this crucial operator in the context $f=2$. A rapid look at Eq. (9) can convince us that in order to commute with the interacting part of the BH Hamiltonian, $T$ has to be a function of the sum of the two scaling operators of $sl(2 ; R)$ i.e. $(J_+ + J_-)$. This function can be specified by asking for the commutation of the diagonal part $J_0^2$ with it on the basis. For instance, if $n=3$, the function $c (J_+ + J_-)^3-7 c (J_+ + J_-)$ does commute with $H_{BH}$ when the irrep $D^{(\frac{3}{2})}$ is under consideration. The constant $c$ is then fixed according to f.i. the requirement $T |3,0 > = |0,3 >$ which gives $c=\frac{1}{6}$. We thus have
\begin{eqnarray}
T& = &\frac{1}{6} ( (J_+ + J_-)^3 - 7 (J_+ + J_-) ) \nonumber \\
& = &x^3 + x^2 (1-x^2) \frac{d}{dx} + \frac{1}{2} x (1-x^2)^2 \frac{d^2}{dx^2} + \frac{1}{6} (1-x^2)^3 \frac{d^3}{dx^3}
\end{eqnarray}
according to Eq. (8). This result is generalized to
\be
T = \sum_{j=0}^n \frac{1}{j!} \; x^{n-j} (1-x^2)^j \frac{d^j}{dx^j}
\ee
for an arbitrary $n$.
\par
Let us now consider the case $f=3$. The algebra $sl(3 ; R)$ is the one subtended by the corresponding BH model. It is generated by 8 operators given in terms of annihilation and creation operators following the original model or, equivalently,  in terms of differential operators :
\be
a^{\dagger}_2 a_1 \sim \partial_{x_1} \; , \; a^{\dagger}_1 a_2 \sim -x_1^2 \partial_{x_1}-x_1 x_2  \partial_{x_2}+n x_1 \; , \; a^{\dagger}_2 a_2-a^{\dagger}_1 a_1 \sim -2 x_1 \partial_{x_1}-x_2 \partial_{x_2}+n,
\ee
\be
a^{\dagger}_2 a_3 \sim -\partial_{x_2} \; , \; a^{\dagger}_3 a_2 \sim x_2^2 \partial_{x_2}+x_1 x_2  \partial_{x_1}-n x_2 \; , \; a^{\dagger}_3 a_3-a^{\dagger}_2 a_2 \sim  x_1 \partial_{x_1}+2x_2 \partial_{x_2}-n,
\ee
\be
a^{\dagger}_3 a_1 \sim -x_2 \partial_{x_1} \; , \; a^{\dagger}_1 a_3 \sim -x_1 \partial_{x_2}.
\ee
Both forms are two different realizations of the same irrep $D(0,n)$ of $sl(3 ; R)$ as clear from the dimension $\frac{1}{2} (n+1) (n+2)$ of the basis these operators act on as well as the eigenvalues of the "hypercharge" and "$T_3$" operators \cite{Greiner} i.e.
\be
Y= \frac{1}{3} (2 a^{\dagger}_3 a_3 - a^{\dagger}_1 a_1 - a^{\dagger}_2 a_2) \sim x_2  \partial_{x_2}-\frac{n}{3}
\ee
and
\be
T_3 = \frac{1}{2} (a^{\dagger}_2 a_2 - a^{\dagger}_1 a_1) \sim - x_1 \partial_{x_1}-\frac{1}{2} x_2  \partial_{x_2} + \frac{n}{2},
\ee  
respectively. The BH model can thus be written as in Eq. (1) or with the differential realization 
\begin{eqnarray}
H_{BH} &=&- \gamma (x_1^2 \partial^2_{x_1}+x_2^2 \partial^2_{x_2}+x_1 x_2 \partial_{x_1}\partial_{x_2})+(x_1^2-1-x_1 x_2 +\gamma(n-1)x_1+x_2) \partial_{x_1} \nonumber \\
&&-(x_2^2-1-x_1 x_2 -\gamma (n-1) x_2-x_1) \partial_{x_2}-n x_1+n x_2 -\frac{\gamma}{2} n (n-1).
\end{eqnarray}
 The equivalence of the bases is
\be
| n - k - j, k , j > \sim x_1^{n-k-j} x_2^j, \; \; j=0, 1, ..., n-k, \; k=0, 1, ..., n.
\ee
The embedding of the cases $f=2$ and $f=3$ is thus clear. It is not possible here to convert the BH Hamiltonian into a Schr\" odinger form but it is rather straightforward to determine its eigenvalues and eigenfunctions either on the bosonic or on the differential forms. 
\par
Here also, it is possible to express the operator $T$ in terms of differential operators. However, it has to be done by hand and no general expression is available. For instance, if $n=1$ we have
\be
T = -( x_1 + x_2 ) (\partial_{x_1} + \partial_{x_2}) +1
\ee
or in terms of bosonic operators
\be
T = a^{\dagger}_3 a_1 + a^{\dagger}_1 a_3 + a^{\dagger}_2 a_2
\ee
while if $n=2$, we obtain
\begin{eqnarray}
T = && \frac{1}{2} (1+x_1 x_2)^2 \partial^2_{x_1}+\frac{1}{2} (x_1-x_2^2)^2 \partial^2_{x_2}+(1+x_1 x_2)(x_2^2-x_1)\partial_{x_1}\partial_{x_2} \nonumber \\
&&-x_2 (1+x_1 x_2) \partial_{x_1} + x_2 (x_1-x_2^2) \partial_{x_2}+x_2^2
\end{eqnarray}
which is not a polynomial of the operator (23).
\par
The generalization to the cases of higher $f$ does not present any difficulty except for its heaviness, technically speaking.
\section{The QES model}
The Hamiltonian which we consider now is given by
\be
H = H_{BH} + H_{\lambda}
\ee
where the second piece $H_{\lambda}$ is chosen according to
\be
H_{\lambda} = \lambda\sum_{j=1}^f\Bigl(a_j^{\dagger}(N-2) + (N-2)a_j\Bigr),
\ee
with the total particle number operator $N$
\be
N = \sum_{j=1}^f a_j^{\dagger}a_j.
\ee
If the operator $H_{BH}$ preserves separately any subspace ${\cal V}_n$ of the Hilbert space
, the piece $H_{\lambda}$ preserves
only the subspace ${\cal V}_0 \oplus {\cal V}_1\oplus {\cal V}_2$ .
In this sense, the full Hamiltonian $H$ is said quasi exactly
solvable \cite{turbiner} since it preserves a finite-dimensional
subspace of the full Hilbert space.  
Accordingly, the Hamiltonian can be diagonalized algebraically
on the subspace. We call this operator the QES BH Hamiltonian.
Note that the occupation
 number operator $N$ does not commute with $H$ while $T$ does.
\section{Group theoretical approach of the QES model}
The QES BH model (26) implies that we add now the bosonic annihilation and creation operators to their bilinear products previously considered. If we ask for commutation relations only, the algebra will not close, that is why we are naturally led to associate a $Z_2$-grading to the operators involved in Eq. (26). Hence the operators $a^{\dagger}_j$ and $a_j$ have an odd parity (and thus obey anticommutation relations) while their bilinear products are even (and have to satisfy commutation relations). So, to the previous $(f^2-1)$ even operators  $a^{\dagger}_j a_k$, we add now $\frac{1}{2} f(f+1)$ even ones given by  $a^{\dagger}_j a^{\dagger}_k$, their conjugates as well as the number operator which is not a invariant anymore. We thus obtain $(2 f^2+f)$ even operators. We complete the structure with the $2f$ annihilation and creation odd operators. These operators generate the Lie orthosymplectic superalgebra $osp(1/2f ; R)$ as is well known \cite{FSS}.
\par
The dimension of the $osp(1/2f ; R)$ irreps involved by the QES BH model can be determined according to the counting of the states : 1 for ${\cal V}_0$, $f$ for ${\cal V}_1$ and $\frac{1}{2} f (f+1)$ for ${\cal V}_2$, giving a total of $\frac{1}{2} (f+1) (f+2)$. The situation is thus different from the one encountered in the usual BH model. In the BH model, the number $f$ of sites determines the Lie algebra while the number $n$ of quanta fixes the involved irrep. In the QES version, the number of quanta is fixed at the start and the number of sites selects both the Lie superalgebra and its irrep.
\par
Let us once again put our attention on two significant cases.
\par
If $f=1$, the fundamental irrep, with $-\frac{1}{16}$ as the eigenvalue of the Casimir operator \cite{FSS}, of $osp(1/2 ; R)$ is under consideration. It can be realized through bosonic operators as in Eq. (26) or via matricial differential operators. In what concerns the odd operators, we have \cite{ShKh}
\be
a = \sqrt{2}
\left( 
\begin{array}{cc}
0 & \frac{d}{dx} \\
1 & 0 
\end{array}
\right) \; , \; 
a^{\dagger} = \sqrt{2} \left(
\begin{array}{cc}
0 & x \frac{d}{dx} + \frac{1}{2} \\
x &0
\end{array}
\right)
\ee
which implies
\be
 H =  \left(
\begin{array}{ll}
-\gamma (2x^2 \frac{d^2}{dx^2} +3 x \frac{d}{dx})-4 x \frac{d}{dx} -2 & \sqrt{2} \lambda (2x(x+1) \frac{d^2}{dx^2}+(x-1) \frac{d}{dx}-1) \\
 \sqrt{2} \lambda (2x(x+1) \frac{d}{dx}-x-2 & -\gamma (2x^2 \frac{d^2}{dx^2} + x \frac{d}{dx})-4 x \frac{d}{dx}
\end{array}
\right)
\ee
The equivalence at the level of the states is
\be
| 0 > \sim \left( 
\begin{array}{c}
0 \\
1
\end{array}
\right)  , 
| 1 > \sim \left( 
\begin{array}{c}
1 \\
0
\end{array}
\right)  ,
| 2 > \sim \left( 
\begin{array}{c}
0 \\
x
\end{array}
\right).
\ee
The energies are determined straightforwardly whatever the 
form of $H$ is. They are the solutions of the equation
\be
E^3+(\gamma +6) E^2+(2 \gamma +8 -6 \lambda^2) E 
-16 \lambda^2 - 4 \lambda^2 \gamma = 0.
\ee
The translation operator $T$ still commutes with $H$ since it commutes with both $a$ and $a^{\dagger}$. It simply reduces to the identity operator.
\par
Let us now turn to the $f=2$ context.
\par
The six-dimensional $osp(1/4 ; R)$ irrep is under consideration here. It can be realized with differential operators of two variables \cite{BD}. Nevertheless, we choose here a differential realization in terms of one variable only since we want to compare the $\lambda$ contribution with respect to the original model. In this aim, we consider the Hamiltonian (10) restricted (due to the QES features) to the cases $n=0, 1, 2$. It reads
\be
 H_{BH} = \left(
\begin{array}{lll}
0 & 0 & 0 \\
0 & 2(x^2-1) \frac{d}{dx} -2x & 0 \\
0 & 0 & 2(x^2-1) \frac{d}{dx}-4x-\gamma (x^2 \frac{d^2}{dx^2}-x \frac{d}{dx}+1)
\end{array}
\right)
\ee
At the level of the states we have
\begin{eqnarray}
| 0 , 0 > \sim \left( 
\begin{array}{c}
1 \\
0 \\
0
\end{array}
\right)  , 
| 1 , 0 > \sim \left( 
\begin{array}{c}
0 \\
1 \\
0
\end{array}
\right)  ,
| 0 , 1 > \sim \left( 
\begin{array}{c}
0 \\
x \\
0
\end{array}
\right)  , \nonumber \\
| 2 , 0 > \sim \left( 
\begin{array}{c}
0 \\
0 \\
1
\end{array}
\right) ,
| 0 , 2 > \sim \left( 
\begin{array}{c}
0 \\
0 \\
x^2
\end{array}
\right)  ,
| 1 , 1 > \sim \sqrt{2} \left( 
\begin{array}{c}
0 \\
0 \\
x
\end{array}
\right)  .
\end{eqnarray}
Knowing these states as well as the action of the annihilators and the creators on them f.i.
\begin{eqnarray}
a_1 | 0 , 0 > = 0 \; , \; a_1 |1 , 0 > = | 0 , 0 > \; , \; a_1 | 0 , 1 > = 0 \; , \nonumber \\
a_1 | 2 , 0 > = \sqrt{2} | 1 , 0 > \; , \; a_1 | 0 , 2 > = 0 \; , \; a_1 | 1 , 1 > = | 0 , 1 >
\end{eqnarray}
we can determine the realization (on this basis) of the odd generators of $osp(1/4; R)$. We obtain
\be
a_1=
\left(
\begin{array}{ccc}
0&-x \frac{d}{dx} + 1 &0 \\
0 &0 &-\frac{1}{\sqrt{2}}x \frac{d}{dx} + \sqrt{2}\\
0 &0 &0
\end{array}
\right) \; , \; a_2=
\left(
\begin{array}{ccc}
0& \frac{d}{dx}  &0 \\
0 &0 &\frac{1}{\sqrt{2}} \frac{d}{dx} \\
0 &0 &0
\end{array}
\right)
\ee
and
\be
a^{\dagger}_1=
\left(
\begin{array}{ccc}
0&0 &0 \\
1 &0 &0\\
0 &\sqrt{2} &0
\end{array}
\right) \; , \; a^{\dagger}_2=
\left(
\begin{array}{ccc}
0&0  &0 \\
x &0 &0 \\
0 &\sqrt{2} x &0
\end{array}
\right).
\ee
This implies (as expected)
\be
N=
\left(
\begin{array}{ccc}
0&0  &0 \\
0 &1 &0 \\
0 &0 &2
\end{array}
\right).
\ee
The QES BH Hamiltonian is then the sum of the BH Hamiltonian (33) and the QES part
\be
H_{\lambda} = 
\left(
\begin{array}{ccc}
0 &2 \lambda ((x-1) \frac{d}{dx}-1) &0 \\
-2 \lambda (x+1) &0 &\lambda ( \frac{1}{\sqrt{2}} (x-1) \frac{d}{dx}-\sqrt{2}) \\
0 &-\sqrt{2} \lambda (x+1)  &0
\end{array}
\right).
\ee
If $\lambda = 0$, the respective eigenstates and eigenvalues are
\be
| 0 , 0 > \leftrightarrow E = 0,
\ee
\be
| 1 , 0 > \pm | 0 , 1 > \leftrightarrow E = \mp 2,
\ee
\begin{eqnarray}
&&| 2 , 0 > - | 0 , 2 > \leftrightarrow E = -\gamma \; , \; \nonumber \\
&&| 2 , 0 > + | 0 , 2 > -\frac{1}{2 \sqrt{2}} (E+\gamma) | 1 , 1 > 
\leftrightarrow 
E = - \frac{\gamma}{2} \pm \frac{1}{2} \sqrt{\gamma^2+64}.
\end{eqnarray}
If $\lambda \neq 0$, the three sectors are mixed and this gives
\be
\lambda \sqrt{2} (| 1 , 0 > - | 0 , 1 >)
+  (2-E) (| 2 , 0 > - | 0 , 2 >)  \ee
for the eigenvalues
\be
E = 1 - \frac{\gamma}{2} \pm \frac{1}{2} 
\sqrt{(\gamma+2)^2+8\lambda^2},
\ee
while the states
\be
c_1 | 0 , 0 > + c_2 (| 1 , 0 > + | 0 , 1 >) + c_3 (| 2 , 0 > + | 0, 2 >) + c_4 | 1 , 1 >
\ee
with
\be
c_1 = 4 \sqrt{2} (E+\gamma -4) \lambda^2
\ee
\be
c_2 = -\sqrt{2} E (E+\gamma -4) \lambda
\ee
\be
c_3 = -(4E^2+(8-2\lambda^2)E-32)
\ee
\be
c_4 = \sqrt{2} (E^3+(\gamma + 2) E^2
+(2\gamma -10\lambda^2) E-8 \lambda^2 \gamma
\ee
are associated with the solutions of
\be
E^4 + (\gamma + 2) E^3 
+ (2\gamma - 16 -12 \lambda^2) E^2 + 
(16 \lambda^2 - 10\gamma \lambda^2 - 32) E + 128 \lambda^2 = 0.
\ee
\section{Diagonalisation of the Hamiltonian}
In order to achieve this diagonalisation,
we will now use a suitable basis of the subspaces ${\cal V}_n$, $n=0,1,2$.
It turns out to be helpfull to write down the matrix elements
$H_{ij}$ in a simple way for generic values of the 
number $f$ of oscillators.
 Along with \cite{scott, eilbeck}, we define
the discrete momentum
\be
\label{nu}
k\equiv \frac{2\pi\nu}{f} \ \
{\rm for} \ \nu = \frac{f-1}{2}, \frac{f-3}{2}, \dots, -\frac{f-1}{2}
\ee
The vector containing no quanta is noted
\be
\vert 0 \rangle = [00\dots0]
\ee
The vectors containing a single quantum are treated
by mean of the $f$ vectors of the form
\be
\vert \psi_1(k)\rangle = \frac{1}{\sqrt f}\sum_{j=1}^f(e^{ik}T)^{j-1}[100\dots 0]
\ee
and the
$\frac{f(f+1)}{2}$ vectors containing two quanta are
of the form $\vert\psi_{2,b}(k)\rangle$ 
\begin{eqnarray}
& \vert \psi_{2,1}(k)\rangle&= \frac{1}{\sqrt f}\sum_{j=1}^f(e^{ik}T)^{j-1}[200\dots 0]\nonumber \\
&\vert\psi_{2,2}(k)\rangle & = \frac{1}{\sqrt f}\sum_{j=1}^f(e^{ik}T)^{j-1}[110\dots 0]\nonumber \\
&\vert\psi_{2,3}(k)\rangle & = \frac{1}{\sqrt f}\sum_{j=1}^f(e^{ik}T)^{j-1}[1010\dots 0]\nonumber \\
&\dots& \nonumber \\
&\vert\psi_{2,b}(k)\rangle & = \frac{1}{\sqrt f}\sum_{j=1}^f(e^{ik}T)^{j-1}[10\dots 010\dots00]
\end{eqnarray}
where the index $b$ takes values $b = 1,2,\dots, \frac{f+1}{2}$
and  it is understood that there are $b-2$ "$0$" between
the two "$1$" in the different vectors (apart from the case $b=1$).

The restriction $H_R$ of $H$ to the invariant vector-space
${\cal V}_0 \oplus {\cal V}_1\oplus {\cal V}_2$
leads to a matrix form
%%%%ICI NATH debut
\be
H_R= \left(
\begin{array}{lll}
0 & H_{01} & 0 \\
H_{01}^{\dagger} & H_{11} & H_{12} \\
0 & H_{12}^{\dagger} & H_{22}
\end{array}
\right)
\ee
where
$$
(H_{01})_\mu = -2 \lambda \sqrt{f} \delta_{\mu 0},
$$
$$
(H_{11})_{\mu \nu} = -2 \delta_{\mu \nu} \cos{\frac{2 \pi \mu}{f}}.
$$
\subsection{$f$ odd}
Then, if $f$ is odd,
\be
(H_{12})_{\mu b, \nu}=
-\sqrt{2} \lambda \delta_{\mu \nu} \delta_{1b}
-\lambda \sum_{j=1}^{\frac{f+1}{2}-1}
(1+\exp{\frac{2 j i \pi \nu}{f}})\delta_{\mu \nu} 
\delta_{j+1 b}, b=1,2, ..., \frac{f+1}{2}
\ee
and
\be
H_{22} = -\gamma -4 \; {\rm if} \; f=1
\ee
while
\be H_{22}=
\left(
\begin{array}{ccc}
-\gamma I_f &-\sqrt{2}\; q^* I_f &O_{\frac{1}{2} f^2-\frac{3}{2}f} \\
-\sqrt{2}\; q I_f &O_{\frac{1}{2} f^2-\frac{3}{2}f} &-q^* I_f \\
O_{\frac{1}{2} f^2-\frac{3}{2}f} &-q I_f &-p \; I_f
\end{array}
\right)
\ee
for other values of odd $f$ with
$$q=1+\exp{\frac{2 i \pi \nu}{f}} \; , 
\; p = \exp{\frac{ i (f+1) \pi \nu}{f}} +  
\exp{\frac{ i (f-1) \pi \nu}{f}} \; , \; 
\nu = \frac{f-1}{2}, \frac{f-3}{2}, ..., -\frac{f-1}{2}$$
while $I_f$ stands for the identity matrix of dimension 
$f$ and $O_{\frac{1}{2} f^2-\frac{3}{2}f}$ for the null 
matrix of dimension $(\frac{1}{2} f^2-\frac{3}{2}f)$.\\
The Hamiltonian further splits into 1 block of dimension 
$\frac{f+5}{2}$ and $(f-1)$ blocks of dimension $\frac{f+3}{2}$.
\par
The different
blocks can still be labelled by the momentum $k$ which, along with 
\cite{scott} renders the classification of the eigenvalues in 
function of $k$ possible. Note that this contrasts with the pure
BH case \cite{scott} where there are $f$ blocks of dimension $(f+1)/2$.
The occurence of the supplementary dimensions in the blocks
is due to the fact that
the invariant subspace contains the vectors with zero and one quantum
as well. These vectors  naturally mix with the ones of ${\cal V}_2$
in the diagonalisation.
\subsection{$f$ even}
\par
Now, if $f$ is even,
\be
(H_{12})_{\mu b, \nu}=-\sqrt{2} 
\lambda \delta_{\mu \nu} \delta_{1b}-
\sqrt{2} \lambda \delta_{\mu \nu} \delta_{\frac{f}{2}+1 b}
-\lambda \sum_{j=1}^{\frac{f}{2}-1}(1+\exp{\frac{2 j i \pi \nu}{f}})
\delta_{\mu \nu} \delta_{j+1 b}, b=1,2, ..., \frac{f}{2}+1
\ee
the second term appearing when $\mu$ is even only. We also have
\be
H_{22} = \left(
\begin{array}{ccc}
-\gamma &0 &0 \\
0 & -\gamma &-4 \\
0 &-4 &0
\end{array}
\right)
 \; {\rm if} \; f=2
 \ee
while
\be 
H_{22}=
\left(
\begin{array}{cccc}
-\gamma I_f &-\sqrt{2} q^* I_f &O_{\frac{1}{2}f^2-2f} &O_{(\frac{1}{2}f^2-2f) \; * \; (\frac{f}{2})} \\
-\sqrt{2} q I_f &O_f &-q^* I_{\frac{1}{2}f^2-2f} &O_{(\frac{1}{2}f^2-2f) \; * \; (\frac{f}{2})} \\
O_{ \frac{1}{2}f^2-2f} &-q   I_{\frac{1}{2}f^2-2f} &O_{\frac{1}{2}f^2-2f} &B \\
O_{(\frac{f}{2}) \; * \; (\frac{1}{2}f^2-2f)} &O_{(\frac{f}{2}) \; * \; (\frac{1}{2}f^2-2f)} &B^{\dagger} &O_{\frac{f}{2}}
\end{array}
\right)
\ee
for other values of even $f$ with
$$ \nu = \frac{f}{2}, \frac{f}{2}-1, ..., -\frac{f}{2}+1.$$ 
In this last matrix, the notation $O_{m \; *\;  n}$ 
stands for the null matrix with $m$ rows and $n$ 
columns while the matrix $B$ is equal 
to $-\sqrt{2} q^* \sum_{j=1}^{\frac{f}{2}} e_{2j-1,j}$ 
with $e_{m,n}$ standing for a matrix where we can find 
zeroes everywhere except at the intersection of the $m^{th}$ 
row and the $n^{th}$ column where a 1 is.\\
The Hamiltonian further splits into 
1 block of dimension $\frac{f+6}{2}$, $\frac{f}{2}$ 
blocks of dimension $\frac{f+2}{2}$ and $\frac{f-2}{2}$ 
blocks of dimension  $\frac{f+4}{2}$.
%%%%ICI   NATH FIN

\section{Examples}
We have studied the spectrum of the matrix $H_R$ for a few values
of $f$. As said above the different eigenvalues can still be
labelled by $k$ and it is possible to plot them on a diagram
with the momentum $k$ set on the horizontal axis.
Drawing such graphics for different $f$ and fixed values
of $\lambda$,$\gamma$ leads the same pattern as the one of Fig.1 of
\cite{scott}.
The eigenvalues of $H_R$ are represented on Fig. 1 for
$\lambda \in [0,0.5]$ and for
$\gamma=3,f=3$.  The label $n$ refers to the number of quanta
defined naturally in the $\lambda = 0$ limit.
The two lowest lines
(labelled $n=2$, $k=\pm 1$) represent the evolution
of the soliton band's eigenvalues for the QES-extended BH model.
It is clearly seen that these eigenvalues stay below the
others for a large interval of the new coupling constant $\lambda$.
A similar analysis in the cases $f=5,7$ indicates the same phenomenon.

In order to illustrate the evolution of the algebraic part of the
spectrum and the splitting of the soliton band in the QES model,
we superpose on Fig. 2 the eigenvalues available in
the case  $\gamma=3,f=7$ for two values of $\lambda$.
Here the eigenvalues are plotted as functions of $\nu$;
the symmetric part (i.e. for $\nu \rightarrow -\nu$)
has, of course, to be supplemented.
The solid-black (resp. dashed-red) 
lines join eigenvalues corresponding
to $\lambda = 0$ (resp. $\lambda = 0.5$). The various eigenvalues
are represented by triangle, square and bullet symbols according
to the fact that they are related to
the $n=0$, $n=1$ and $n=2$ sectors occuring in the $\lambda = 0$ limit.
The lower curve represents the soliton band.
The picture clearly suggests the
persistence of this band for $\lambda > 0$.
 All other eigenvalues turn out
to be located inside an  envelope.
To finish, we present some detailed calculations
of the eigenvalues and of the eigenvectors for $f=1,2,3,4$

\begin{figure}
\epsfysize=22cm
\epsffile{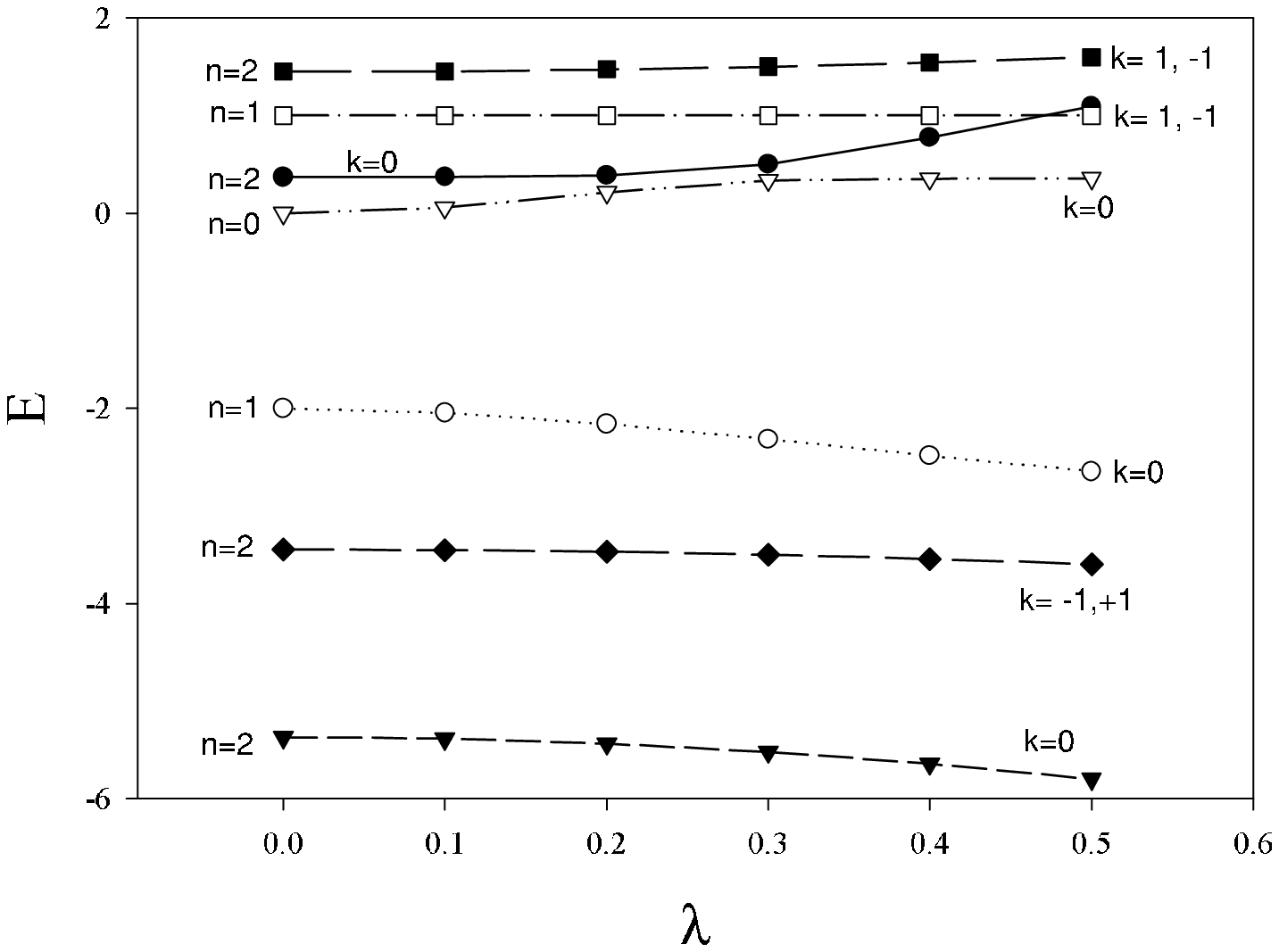}
\vskip -3cm
\caption{\label{Fig.1} The energy eigenvalues
corresponding to the case $f=3,\gamma=3$ are plotted
as functions of the coupling constant $\lambda$.
}
\end{figure}
\begin{figure}
\epsfysize=22cm
\epsffile{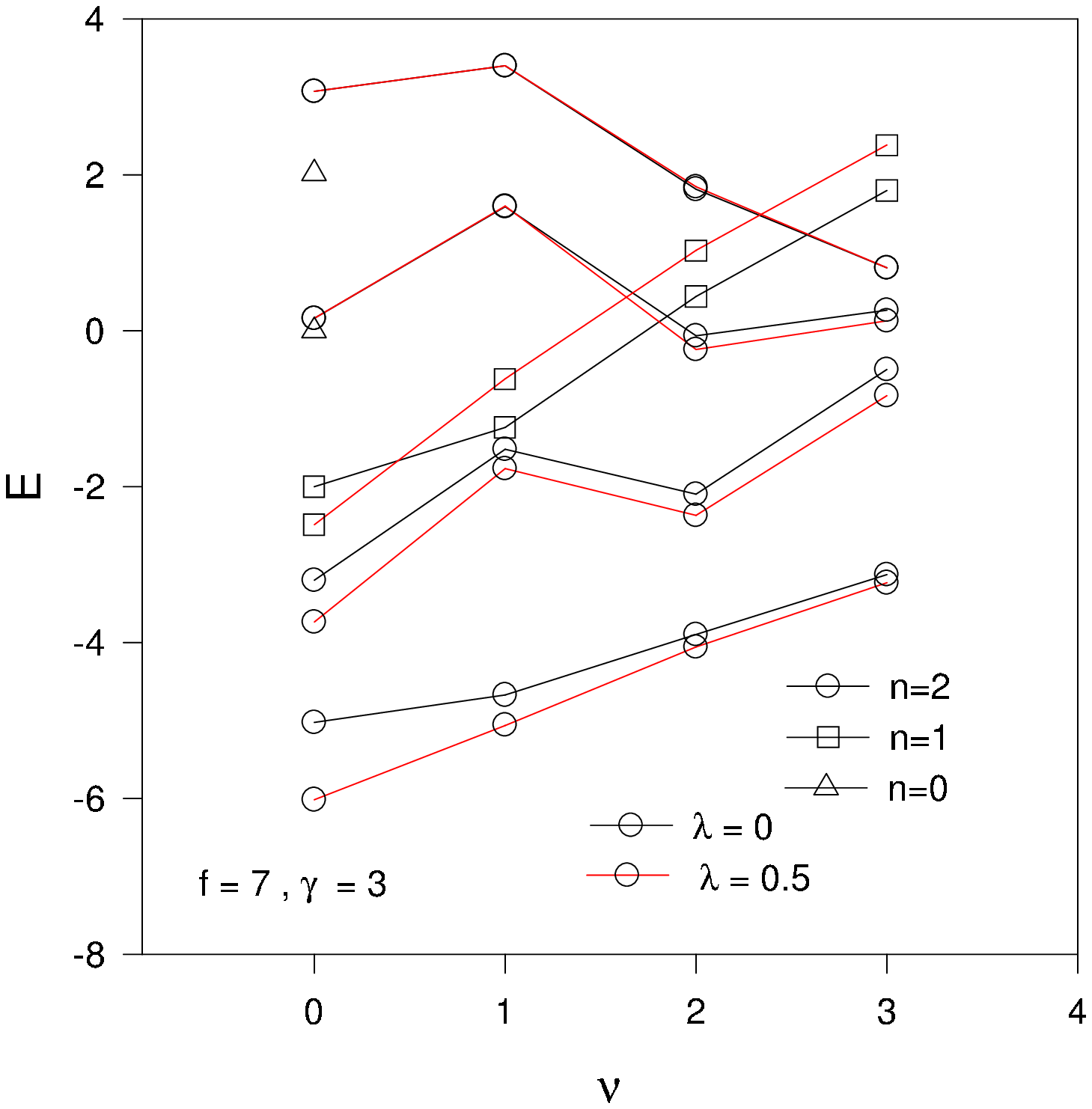}
\vskip -3cm
\caption{\label{Fig.2} The energy eigenvalues
corresponding to the case $f=7,\gamma=3$ are plotted
as functions of the momentum $\nu$ for $\lambda=0.0$ and  $\lambda=0.5$.
}
\end{figure}

\subsection{The cases f=1 and 3}
%%%%%%%%%%%%EXEMPLE NATH DEBUT
%%%Some examples : \vspace{3mm} \\
If $f=1$, the energies satisfy the equation
$$ E^3+(\gamma + 6)E^2+(2\gamma +8-6\lambda^2) E-16 \lambda^2-4 \gamma \lambda^2 = 0$$
which gives for $\gamma =3$ \vspace{3mm} \\
\begin{tabular}{|l|lll|}
\hline
$\lambda$ & $E_1$ & $E_2$ & $E_3$ \\
\hline
0.0 & -7.000 & -2.000 & 0.000 \\
0.1 & -7.004 & -2.016 & 0.020 \\
0.2 & -7.016 & -2.061 & 0.077 \\
0.3 & -7.036 & -2.132 & 0.168 \\
0.4 & -7.064 & -2.221 & 0.286 \\
0.5 & -7.101 & -2.323 & 0.424 \\
\hline
\end{tabular} \vspace{5mm} \\
for the eigenstates 
$$
2 \sqrt{2} \lambda^2 |0> - \sqrt{2} \lambda E |1> + (E^2+2E-4\lambda^2) |2>.
$$
If $f=3$, we have a split : one block of dimension 4 and two others of dimension 3. Concerning the first block, the related equation is
$$
E^4+(\gamma +4)E^3+(4 \gamma -4-18 \lambda^2)E^2+(4\gamma -16-12\lambda^2-16 \gamma \lambda^2)E+96 \lambda^2-24 \gamma \lambda^2=0
$$
which gives for $\gamma =3$ \vspace{3mm} \\
\begin{tabular}{|l|llll|}
\hline
$\lambda$ & $E_1$ & $E_2$ & $E_3$ &$E_4$\\
\hline
0.0 & -5.372 & -2.000 & 0.000 &0.372\\
0.1 & -5.389 & -2.043 & 0.058 &0.374 \\
0.2 & -5.439 & -2.159 & 0.212 &0.386 \\
0.3 & -5.524 & -2.314 & 0.336 &0.503 \\
0.4 & -5.645 & -2.484 & 0.353 &0.776 \\
0.5 & -5.801 & -2.649 & 0.357 &1.094 \\
\hline
\end{tabular} \vspace{5mm}\\
for the eigenstates 
\begin{eqnarray}
&&4 \sqrt{6} (E+1) \lambda^2 |0>-2 \sqrt{2} E (E+1) \lambda |\psi_1 (0)>+(-4E^2+(4\lambda^2-8)E+48 \lambda^2) |\psi_{2,1}(0)> \nonumber \\
&&+\sqrt{2} (E^3+5E^2+(6-14 \lambda^2)E-36 \lambda^2) |\psi_{2,2}(0)>. \nonumber
\end{eqnarray}
Concerning the two other blocks, we have the same equation
$$
E^3+(\gamma -2)E^2-(3\lambda^2 +2\gamma +1) E+2+\gamma+6 \lambda^2- \gamma \lambda^2 =0
$$
whose solutions are
$$E=1 \; , \; E=-1 \pm \sqrt{3(2+\lambda^2)}$$
or for $\gamma =3$
\vspace{3mm} \\
\begin{tabular}{|l|lll|}
\hline
$\lambda$ & $E_1$ & $E_2$ & $E_3$\\
\hline
0.0 & -3.450 & 1.000 & 1.450 \\
0.1 & -3.456 & 1.000 & 1.456  \\
0.2 & -3.474 & 1.000 & 1.474  \\
0.3 & -3.504 & 1.000 & 1.504  \\
0.4 & -3.546 & 1.000 & 1.546  \\
0.5 & -3.598 & 1.000 & 1.598  \\
\hline
\end{tabular} \vspace{5mm}\\
The eigenstates are
$$
| \psi_1 (\pm \frac{2\pi}{3}) > - \frac{\lambda}{\sqrt{2}} | \psi_{2,1} (\pm \frac{2\pi}{3}) > + \frac{\lambda}{2}(1 \pm i \sqrt{3}) | \psi_{2,2} (\pm \frac{2\pi}{3}) >$$
for $E=1$ and 
$$
-3(1\mp i \sqrt{3}) \lambda | \psi_1 (\pm \frac{2\pi}{3}) >+\sqrt{2} (1\mp i \sqrt{3})(E-2) | \psi_{2,1} (\pm \frac{2\pi}{3}) >+2(E+1) | \psi_{2,2} (\pm \frac{2\pi}{3}) >
$$
for the other energies. As is clear from above, we observe a twofold degeneracy : the degenerated states are complex conjugated.
%%%%%%%%%%%%%%%%%%%%%%%%%%%%%%%%%%%%%%%%%%%%%%%%%%%%%%%%%%%%%%%%%%%%%%
\par 
\subsection{The cases f=2 and 4}
Some examples : \vspace{3mm} \\
If $f=2$, we have one block of dimension 4 and another one of dimension 2. The energies of the first block satisfy the equation
$$ E^4+(\gamma + 2)E^3+(2\gamma -16-12\lambda^2) E^2+(-32+16\lambda^2-10 \gamma \lambda^2)E+128 \lambda^2 = 0$$
which gives for $\gamma =3$ \vspace{3mm} \\
\begin{tabular}{|l|llll|}
\hline
$\lambda$ & $E_1$ & $E_2$ & $E_3$ &$E_4$\\
\hline
0.0 & -5.772 & -2.000 & 0.000 &2.772\\
0.1 & -5.782 & -2.029 & 0.039 &2.772 \\
0.2 & -5.813 & -2.110 & 0.150 &2.773 \\
0.3 & -5.865 & -2.227 & 0.318 &2.775 \\
0.4 & -5.939 & -2.364 & 0.525 &2.777 \\
0.5 & -6.034 & -2.507 & 0.761 &2.780 \\
\hline
\end{tabular} \vspace{5mm} \\
for the eigenstates 
\begin{eqnarray}
&&4(E-1)\lambda^2 |0> - \sqrt{2} E(E-1) \lambda |\psi_1 (0)>-(4 E^2+(8-2\lambda^2)E-32\lambda^2) |\psi_{2,1}(0)> \nonumber \\
&&+(E^3+5E^2+(6-10\lambda^2)E-24\lambda^2) |\psi_{2,2}(0)> \nonumber
\end{eqnarray}
The energies of the second block satisfy the equation
$$
E^2+(\gamma -2)E-2\gamma -2\lambda^2=0
$$
which gives for $\gamma =3$ \vspace{3mm} \\
\begin{tabular}{|l|ll|}
\hline
$\lambda$ & $E_1$ & $E_2$ \\
\hline
0.0 & -3.000 & 2.000 \\
0.1 & -3.004 & 2.004 \\
0.2 & -3.016 & 2.016  \\
0.3 & -3.036 & 2.036  \\
0.4 & -3.063 & 2.063  \\
0.5 & -3.098 & 2.098  \\
\hline
\end{tabular} \vspace{5mm} \\
for the eigenstates 
$$
\sqrt{2} \lambda |\psi_1 (\pi)>+(2-E)  |\psi_{2,1} (\pi)>.
$$

If $f=4$, we still have a split : one block of dimension 5, another one of dimension 4 and two of dimension 3. Concerning the first block, the related equation is
\begin{eqnarray}
&&E^5+(\gamma +2)E^4+(2 \gamma -16-24 \lambda^2)E^3+(-8\gamma -32+32\lambda^2-22 \gamma \lambda^2)E^2 \nonumber \\
&&+(-16 \gamma+256 \lambda^2+16 \gamma \lambda^2) E+128 \gamma \lambda^2=0 \nonumber
\end{eqnarray}
which gives for $\gamma =3$ \vspace{3mm} \\
\begin{tabular}{|l|lllll|}
\hline
$\lambda$ & $E_1$ & $E_2$ & $E_3$ &$E_4$ &$E_5$\\
\hline
0.0 & -5.191 & -2.000 &-1.317 & 0.000 &3.509\\
0.1 & -5.214 & -2.066 &-1.307 & 0.078 &3.509 \\
0.2 & -5.282 & -2.228 &-1.288 & 0.289 &3.509 \\
0.3 & -5.398 & -2.429 &-1.273 & 0.590 &3.510 \\
0.4 & -5.563 & -2.631 &-1.263 & 0.947 &3.510 \\
0.5 & -5.778 & -2.814 &-1.257 & 1.338 &3.511 \\
\hline
\end{tabular} \vspace{5mm}\\
for the eigenstates 
\begin{eqnarray}
&&4 \sqrt{2} (E-4)(E+3) \lambda^2 |0>- \sqrt{2} E (E-4) (E+3)\lambda |\psi_1 (0)>\nonumber \\
&&+(-8E(E+2)-2(-64+E(E-8))\lambda^2) |\psi_{2,1}(0)> +2\sqrt{2}(E+3) (-E(E+2)\nonumber \\
&&+(E+16) \lambda^2) |\psi_{2,2}(0)>\nonumber \\
&&+(128\lambda^2+E(E+2)(-8+E(E+3)-22\lambda^2))  |\psi_{2,3}(0)>. \nonumber
\end{eqnarray}
Concerning the block of dimension 4, we have
$$
E(E^3+(\gamma -2)E^2-(4\lambda^2+2\gamma)E-2\gamma \lambda^2)=0
$$
which gives for $\gamma =3$ \vspace{3mm} \\
\begin{tabular}{|l|llll|}
\hline
$\lambda$ & $E_1$ & $E_2$ & $E_3$ &$E_4$ \\
\hline
0.0 & -3.000 & 0.000  &0.000 &2.000\\
0.1 & -3.004 & -0.010 &0.000 &2.014 \\
0.2 & -3.016 & -0.039 &0.000 &2.055 \\
0.3 & -3.036 & -0.084 &0.000 &2.120 \\
0.4 & -3.065 & -0.142 &0.000 &2.206 \\
0.5 & -3.101 & -0.209 &0.000 &2.311 \\
\hline
\end{tabular} \vspace{5mm}\\
The eigenstate is
$$
|\psi_{2,2}(\pi) >
$$
for the null energy and the other states are
$$
\sqrt{2} (E+3) \lambda |\psi_1 (\pi)>-2 \lambda^2 |\psi_{2,1}(\pi)>-(E^2+E-2\lambda^2-6) |\psi_{2,3}(\pi)>
$$
in what concerns the three other energies.

Concerning the two final blocks, we have the same equation
$$
E^3+\gamma E^2-4(\lambda^2  +1) E+8 \lambda^2-2 \gamma \lambda^2 =0
$$
and for $\gamma =3$
\vspace{3mm} \\
\begin{tabular}{|l|lll|}
\hline
$\lambda$ & $E_1$ & $E_2$ & $E_3$\\
\hline
0.0 & -4.000 & 0.000 & 1.000 \\
0.1 & -4.009 & 0.005 & 1.004  \\
0.2 & -4.036 & 0.020 & 1.016  \\
0.3 & -4.080 & 0.043 & 1.037  \\
0.4 & -4.140 & 0.072 & 1.067  \\
0.5 & -4.215 & 0.107 & 1.107  \\
\hline
\end{tabular} \vspace{5mm}\\
The eigenstates are
$$
(1+E) \lambda |\psi_1 (\pm \frac{\pi}{2})>-\sqrt{2} (\lambda^2-E)|\psi_{2,1} (\pm \frac{\pi}{2})>-\frac{1}{2}(1\pm i)(E^2+3E-2\lambda^2) |\psi_{2,2} (\pm \frac{\pi}{2})>.
$$
As is once again clear from above, we observe a twofold degeneracy : 
the degenerated states are complex conjugated.

%%%%%%%%%%%%%FIN EXEMPLE NATH

\section{Concluding remarks}
The exactly solvable models of the BH type can naturally be generalized
to a family of $f$-body quasi exactly solvable, translation
invariant Hamiltonians.
These generalisations depend on one (or more) new coupling constant(s).
Here we have put the emphasis on a peculiar property of the spectrum
in the generalized models~: when plotted with respect to the discrete
momentum $k$, a line of eigenvalues (one for each values of $k$)
appears splitted from the rest of the spectrum, forming a soliton band.\\
\vskip 2 cm
%%\noindent
{\bf Acknowledgements}
\noindent
Y. B. gratefully ackowledges 
J.C. Eilbeck for discussions and
the organizers of the Symposium
"Topological Solitons and their Applications" held in
Durham (G.B.) in August 2004 for their invitation.
A. N. is supported by a grant of the C.U.D..
%%%\newpage

\end{document}